\documentclass[
aps,
longbibliography,
superscriptaddress,
 amsmath,amssymb,
twocolumn,
]{revtex4-1}

\usepackage{graphicx}
\usepackage{dcolumn}
\usepackage{bm}
\usepackage{upgreek}
\usepackage{color}
\usepackage{hyperref}
\usepackage{amsmath}

\newcommand{\RR}{\right}
\newcommand{\LL}{\left}
\newcommand{\m}{\mathrm}
\newcommand{\dg}{\dagger}
\newcommand{\eref}[1]{Eq.~(\ref{#1})}
\newcommand{\fref}[1]{Fig.~\ref{#1}}
\newcommand{\puoli}{\frac{1}{2}}

\definecolor{gold}{RGB}{215,155,0}
\definecolor{blue}{RGB}{0,0,255}
\definecolor{red}{RGB}{255,0,0}

\usepackage[normalem]{ulem}

\begin{document}


\title{Near-ground state cooling in electromechanics using measurement-based feedback and Josephson parametric amplifier}


\author{Ewa Rej}
\affiliation{QTF Centre of Excellence, Department of Applied Physics, Aalto University, FI-00076 Aalto, Finland}

\author{Richa Cutting}
\affiliation{QTF Centre of Excellence, Department of Applied Physics, Aalto University, FI-00076 Aalto, Finland}
%

%

\author{Debopam Datta}
\affiliation{QTF Centre of Excellence, VTT Technical Research Centre of Finland Ltd, P.O. Box 1000, FI-02044 VTT,  Finland}

\author{Nils Tiencken}
\affiliation{QTF Centre of Excellence, VTT Technical Research Centre of Finland Ltd, P.O. Box 1000, FI-02044 VTT,  Finland}

\author{Joonas Govenius}
\affiliation{QTF Centre of Excellence, VTT Technical Research Centre of Finland Ltd, P.O. Box 1000, FI-02044 VTT,  Finland}

\author{Visa Vesterinen}
\affiliation{QTF Centre of Excellence, VTT Technical Research Centre of Finland Ltd, P.O. Box 1000, FI-02044 VTT,  Finland}

\author{Yulong Liu}
\affiliation{Beijing Academy of Quantum Information Sciences, Beijing 100193, China}

\author{Mika A. Sillanp\"a\"a}
 \email{Mika.Sillanpaa@aalto.fi}
\affiliation{QTF Centre of Excellence, Department of Applied Physics, Aalto University, FI-00076 Aalto, Finland}
%


\begin{abstract}
Feedback-based control of nano- and micromechanical resonators can enable the study of macroscopic quantum phenomena and also sensitive force measurements. Here, we demonstrate the feedback cooling of a low-loss and high-stress macroscopic SiN membrane resonator close to its quantum ground state. We use the microwave optomechanical platform, where the resonator is coupled to a microwave cavity. The experiment utilizes a Josephson travelling wave parametric amplifier, which is nearly quantum-limited in added noise, and is important to mitigate resonator heating due to system noise in the feedback loop. We reach a thermal phonon number as low as 1.6, which is limited primarily by microwave-induced heating. We also discuss the sideband asymmetry observed when a weak microwave tone for independent readout is applied in addition to other tones used for the cooling. The asymmetry can be qualitatively attributed to the quantum-mechanical imbalance between emission and absorption. However, we find that the observed asymmetry is only partially due to this quantum effect. In specific situations, the asymmetry is fully dominated by a cavity Kerr effect under multitone irradiation.

\end{abstract}

\maketitle


\section{\label{sec:level1} INTRODUCTION}

Preparing and controlling mechanical oscillators in the quantum regime of their motion has been an active subject of research. This has provided a platform for observing macroscopic quantum phenomena \cite{Riedinger,Entanglement,Safavi2019Fock,Teufel2021entangle,Chu2023macrosc}, and is showing promise for studying new physics \cite{Marin2013QGR,HammererColl2014,PaternostroColl2014,Vitali2015qgrav,Oosterkamp2016,Taylor2021Adarkmatter,Singh2021DarkMatter}, especially the connection between quantum mechanics and gravity \cite{Marin2016QGR,Mann2019QGR,Gravity2021,Aspelmeyer2021gra,Plenio2022Genhance}. A carefully engineered system of mechanical oscillators also has applications in other types of quantum sensing \cite{Aguiar2011review,Tobar2014QW,Teufel2017accel,Qvarfort2018gravimetry}. It is quite evident that for all these goals, it is beneficial, if not necessary, to cool the motion of the oscillator close to the quantum ground state. Any unnecessary thermal noise energy will add fluctuations, making it more difficult to observe or utilize quantum correlations. For purely sensing applications, thermal noise will mask the oscillator's response to weak forces.

The frequencies of typical mechanical modes in somewhat massive systems are low compared to thermal energy scales, resulting in a thermally excited phonon number greater than one even at deep cryogenic temperatures \cite{RevModPhys.86.1391}. Therefore, there is a need to develop alternative ways to cool the system. Mechanical devices are most typically brought into the quantum domain by interfacing them with cavity optomechanics, in which the radiation pressure forces from the light field are coupled to the oscillator's motion. 

There are two main experimental approaches to cooling in cavity optomechanical systems that are effective. One of them is sideband cooling \cite{Karrai2004,Aspelmeyer2006cool,Schwab2010,Teufel2011b,LaserCooling}, in which the radiation pressure force from the cavity field is used to autonomously dampen the mechanical motion. The second approach is measurement based feedback control of the oscillator \cite{Cohadon1999,Kleckner,Poggio2007FB,Kippenberg2015FB,Rossi_2018}. It is based on measuring the position of the mechanical oscillator and applying a viscous feedback force through a time delayed loop in accordance with the measurement outcome. Roughly speaking, the sideband cooling scheme is more efficient in the resolved-sideband limit, i.e., when the cavity bandwidth is smaller than the mechanical frequency, which results in a time-delayed force from the slowly changing cavity. Feedback cooling is more suitable in the opposite situation, namely the unresolved-sideband limit, where a fast cavity can track the oscillator's position at a high time resolution. The unresolved-sideband situation arises necessarily when resonators with low frequencies, such as 100 Hz ... 10 kHz are used, because cavity linewidths smaller than this become experimentally unfeasible. Cooled resonators in this regime \cite{Abbott_2009} are becoming more and more relevant in basic studies related to gravity \cite{Gravity2021,Pikovsk2024graviton}.

The experimental challenge faced in feedback cooling an oscillator close to the quantum ground state is that the measurement must happen on a time scale faster than the coherence time of the oscillator. This leads, first of all, to the requirements of a relatively low starting temperature and a high quality factor of the oscillator. Second, the added noise in the measurement cannot be too far from a quantum-limited noise level. After the inception of the concept of feedback cooling in cavity optomechanics \cite{Tombesi1998feedback}, it took a long time before the ground state was reached experimentally \cite{Rossi_2018}, as these requirements were difficult to meet. Typically the experiments utilized a cavity optomechanical system that works truly at optical wavelengths, where the system noise is naturally small, but the setup is not naturally compatible with operation in a dilution refrigerator, which would allow for a low thermal occupancy to begin with. Feedback cooling was recently demonstrated in an electrical realization of cavity optomechanics \cite{2023_feedback}, which works naturally at mK temperatures, but where a larger system noise typical of microwave setups did not allow for near-ground state cooling.


In this work, we combine to the electromechanical feedback setup the necessary prerequisites for ground-state cooling: a high level of mechanical coherence, and a low system noise. Different from earlier work, we introduce a Josephson travelling wave parametric amplifier (JTWPA) \cite{Siddiqi2015Amp}, which reduces the added noise in our system by a factor of five, enabling feedback cooling close to the ground state. As the mechanical device, we use a silicon nitride membrane, which offers one of the best quality factors in the microwave optomechanical setup at cryogenic temperatures \cite{Steele3D,Steele2015HighQ,Nakamura3D,2022_SiNBAE,Schliesser2022SiN}. Additionally, we directly measure the mechanical noise energy through sidebands of an auxiliary weak thermometry tone instead of inferring it from the in-loop noise spectrum as in our earlier work \cite{2023_feedback}. 

Thermometry with a relatively weak tone, as in our current study, can be used to directly demonstrate quantum-mechanical behaviour in optomechanical systems. This derives from the fact that spectral densities of quantum systems are not symmetrical in frequency, which leads to substantial difference in emission and absorption in the quantum oscillator, and asymmetry in the emission spectrum which is not too difficult to observe using currently available techniques \cite{Painter2012ZP,Schwab2014Asymm,Harris2015asym,Kippenberg2020asymm,2021_GrenobleQ}. It was recently pointed out \cite{Kippenberg2019floquet}, however, that such asymmetric emission can be an artefact arising as a combination of multi-tone irradiation and a nonlinear cavity in an optical-regime optomechanical system. We observe and analyze such an artificial sideband asymmetry, which is attributed to fast thermal modulation of the cavity frequency, similar to Ref.~\cite{Kippenberg2019floquet}.

\section{Theoretical discussion}

In this section, we first discuss the scheme of cavity optomechanical measurement, including some subtleties. Then we briefly review how feedback cooling is realized in optomechanics, and how the temperature of the cooled mechanical mode can be measured using an additional thermometry tone. We focus on discussing the thermometry, which becomes not straightforward under the application of multiple tones, which can lead to unwanted couplings between sideband processes.

\subsection{Microwave optomechanical system}

Cavity optomechanics describes the interaction of electromagnetic fields and mechanical motion inside a cavity resonator. Originally, the electromagnetic fields were treated as optical, but more recently, a large amount of research has utilized microwave-frequency fields. These experiments are well-suited operating in dilution refrigerators at temperatures down to 10 mK. We assume a standard cavity optomechanical device, where the electromagnetic cavity with frequency $\omega_c$ is coupled parametrically to a mechanical oscillator of frequency $\omega_m$, such that the oscillator modulates the cavity frequency. The system is described by the Hamiltonian 
\begin{equation}
\label{eq:H0}
\begin{split}
     & H_0 =  \omega_c a^\dg a + \omega_m b^\dg b + g_0 a^\dg a (b^\dg +b) \,.
\end{split}
\end{equation}
Here, the cavity is described by the standard operators $a^\dg$, $a$, and the mechanical oscillator similarly by $b^\dg$ and $b$. The single-photon radiation-pressure coupling is $g_0$, which in our case is determined by the strength of the capacitive coupling between the oscillator and the microwave-frequency cavity.

As typical with mechanical oscillators operating in the high kHz or MHz frequency regime, the thermal phonon number $n_m^T \gg 1$ even at mK temperatures. Although the cavity is in principle thermalized to zero thermal population in the refrigerator, it is often found that the strong driving fields heat up the cavity in microwave experiments, such that a non-zero and significant thermal photon population $n_c^T$ appears.

The energy losses in the system are described as follows: The mechanical oscillator has the intrinsic energy loss rate $\gamma$. For the cavity, we write the loss rate as $\kappa=\kappa_i+\kappa_e$, which consists of internal losses ($\kappa_i$), and external losses ($\kappa_e$), the latter describing coupling to the measurement apparatus. 

In the basic optomechanics measurement, which also forms the starting point of feedback cooling, a single coherent tone of the frequency $\omega_d$ is applied to the device. The detuning from the cavity is $\Delta = \omega_d - \omega_c$. The pumping with the coherent drive induces a photon number $n_c \gg 1$, which results in the effective coupling $G = \sqrt{n_c} g_0$, and the linearized interaction $G (a^\dg + a)  (b^\dg +b)$.

In the reflection measurement, the monitored signal is the electromagnetic field 
\begin{equation}
\begin{split}
& a_\m{out}(\omega) =  \sqrt{\kappa_e} a(\omega) - a_{in,e}(\omega)
     \end{split}
\end{equation}
leaking out from the cavity, which encodes the mechanical information in frequency sidebands of $a_{out}(\omega)$. Here, $a_{in,e}$ is the external input noise, which is assumed to be at the vacuum level.  This signal is amplified with a linear amplifier, characterized by the added number of noise quanta $n_\m{add}$. At this point, the signal can be steered to processing, for example towards generation of the feedback.

In a heterodyne readout, the spectral densities of the mechanical sidebands can be used as a primary thermometer due to the mentioned quantum asymmetry of the sideband weights. Typically, this measurement is carried out by driving the system sequentially at the red or blue sideband, and the relevant mechanical sideband appears at the cavity frequency in the output spectrum  \cite{Painter2012ZP}. The off-resonant sideband is usually suppressed by the cavity susceptibility. With red-sideband driving, $\Delta = -\omega_m$, the relevant sideband is the anti-Stokes process, where photons scatter upwards in frequency. In the blue-detuned case, $\Delta = \omega_m$, the Stokes-scattered sideband is relevant. The spectral weights of the sidebands are denoted by $A^-$ and $A^+$ for the Stokes and anti-Stokes processes, respectively. If the dynamical backaction can be neglected, the weights directly provide the phonon occupation number:
\begin{equation}
\label{eq:nmasym}
\begin{split}
n_m = \frac{A^+}{A^- - A^+} \,.
\end{split}
\end{equation}
For example, at the ground state $n_m = 0$, the anti-Stokes sideband vanishes.

The result in \eref{eq:nmasym} holds, too, if the system is driven at zero detuning $\Delta = 0$, in which case both sidebands are simultaneously observable, moreover, any dynamical backaction does not compromise the interpretation of the sideband weights. We choose a situation close to this in our experiment.

There are several caveats to this basic picture of sideband asymmetry in optomechanics. The simplest deviation from the ideal case is if the detuning $\Delta$ is neither at the red or blue sideband, nor equal to zero. In these cases, the cavity susceptibility distorts the sideband weights. This effect is small in our experiment, but nevertheless  has to be considered in the analysis. Another, more delicate issue is that classical cavity noise adds on top of the zero-point noise in a manner that is qualitatively indistinguishable from the latter, by causing “noise squashing”, i.e., destructive interference for the anti-Stokes sideband \cite{Collin2021uwopto,2021_Grenoble}. The noise emission from the hot cavity, disregarding optomechanics, follows a Lorentzian profile
\begin{equation}
\label{eq:SoutCavNoise}
\begin{split}
S_{\m{out}} =  \frac{\kappa_e \kappa  n_c^T}{(\omega-\omega_c)^2+ ( \frac{\kappa}{2})^2} \,.
\end{split}
\end{equation}
At a relatively small detuning $|\Delta| \ll \kappa$, and operating in the resolved-sideband limit, the contribution of the cavity noise on the sideband asymmetry is negligible, however, with our current experimental parameters, we cannot neglect it.

\subsection{Feedback cooling and detection}

Feedback cooling, also known as cold damping, is a classic example of measurement based control of a system. There is extensive theoretical and experimental literature describing feedback cooling in the standard cavity optomechanical setup. For convenience, we repeat below the main results, which are here extended to cover non-negligible internal cavity losses, and a finite cavity thermal population $n_c^T$, as these are relevant to describe the experiments below. We use the nomenclature introduced in Ref.~\cite{2023_feedback}, which conveniently suits microwave setups where the noise level is described in units of added noise quanta, rather than photon collection efficiency in optics.

\begin{figure}
    \centering
    \includegraphics[width=1\linewidth]{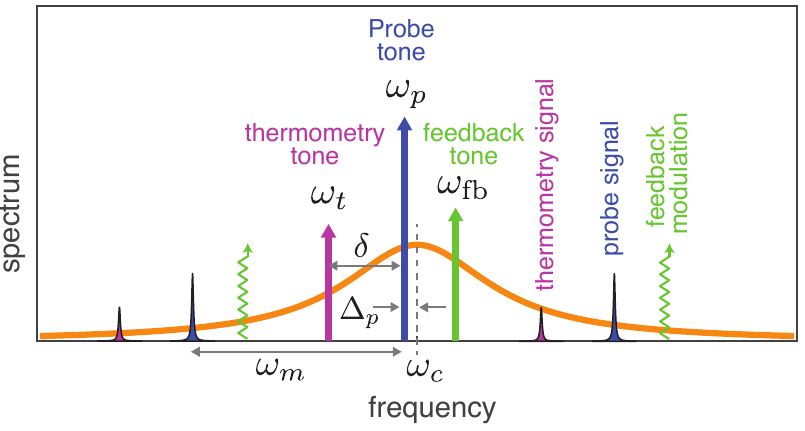}
    \caption{\emph{Frequency scheme of three coherent tones applied to the feedback-controlled microwave optomechanical system.}  The sidebands of the tones are colored similar to the tone itself. For the feedback tone, the sidebands indicate the feedback modulation going into the device, whereas for the other two tones, they represent information coming from the device for detection. The sidebands are labeled only on the right hand side for clarity.}
    \label{fig:tones_setup}
\end{figure}

The system is measured by a strong probe tone at the frequency $\omega_p$, typically at the cavity frequency (see \fref{fig:tones_setup}). This tone is associated with the effective coupling $G_p$. The feedback operation, based on the probe's measurement results, amplifies the mechanical signal by a real-valued and positive amplitude gain $A_0$, as well as applies a time delay such that a phase shift $\phi$ results at the mechanical frequency. By means of phase modulation to an additional feedback tone, the signal is then converted into the feedback force that drives the oscillator.

Depending on the phase shift, the feedback force either counteracts the oscillator's random motion and thus cools it, or amplifies the motion. Immediately below, we assume that the detuning of the probe tone is zero: $\Delta_p = 0$. This implies that the dynamical backaction, i.e., sideband cooling or amplification, is absent. The effective oscillator is characterized by the damping rate 
\begin{equation}
\label{eq:gammaeff}
\begin{split}
\gamma_{\m{eff}} = \gamma + \gamma_{\m{fb}} \frac{ \kappa \sin \phi-2 \omega_m \cos \phi }{\sqrt{\kappa^2+4 \omega_m^2}} \,,
\end{split}
\end{equation}
where the maximum feedback-induced damping at an optimum phase is
\begin{equation}
 \gamma_{\m{fb}} =\frac{4 G_p A_0}{\sqrt{\kappa^2 + 4\omega_m^2} } \,.
\label{eq:gammafb}
\end{equation}

Under optimum feedback conditions, the oscillator experiences cooling down to a thermal phonon number $n_m$ given by
\begin{equation}
\label{eq:nmFB}
\begin{split}
n_m +\puoli= n_T + n_{\m{ba}} + n_{\m{fb}}  \,.
\end{split}
\end{equation}
Here, the thermal and quantum backaction contributions are, including cavity noise, respectively
%
\begin{align}
\label{eq:nT}
 n_T & = \frac{\gamma}{\gamma_{\m{eff}}} \LL(n_m^T + \puoli \RR) \,, \\
\label{eq:nBA}
n_{\m{ba}} & = C_{\m{eff}} \frac{\kappa^2}{\kappa^2 + 4 \omega_m^2} \LL(1 +  2 n_c^T \RR) \,
\end{align}
%
with the effective cooperativity 
\begin{equation}
\label{eq:Ceff}
\begin{split}
C_{\m{eff}} = \frac{4G_p^2}{\kappa \gamma_{\m{eff}}} \,.
\end{split}
\end{equation}
%

The cooling is typically limited by the system noise being injected back to the oscillator, causing heating given by
\begin{equation}
\label{eq:nFB}
\begin{split}
n_{\rm{fb}} = \frac{A_0^2}{2 \kappa_e \gamma_{\m{eff}} } \LL(n_{\m{add}} + \puoli +  \frac{8 \kappa \kappa_e}{  \kappa^2 + 4 \omega_m^2 } n_c^T \RR)  \,.
\end{split}
\end{equation}
The last term in \eref{eq:nBA} in parentheses indicates that cavity noise contributes to the feedback heating by a large weight, because $n_c^T$ enters with a large prefactor. 

In this work, we measure $n_m$ directly by an additional tone, called the thermometry tone, applied at a frequency $\omega_t$, with the corresponding detuning $\Delta_t = \omega_t - \omega_c$, and the effective coupling $G_t$. We also denote as $\delta$ the frequency spacing between the probe tone and the thermometry tone, i.e., $\delta = \omega_p - \omega_t$. The thermometry tone drops sidebands at the frequencies $\omega_t \pm \omega_m$, the energy of which, after calibration, gives $n_m$. The frequency configuration of all the tones applied is shown in \fref{fig:tones_setup}.

\subsection{Cavity Kerr effect}

If several coherent tones are applied to an ideal optomechanical system, the processes they induce can typically be treated as independent, if the corresponding spectra do not overlap. This allows, for example, to carry out independent readout of the mechanical mode in addition to specific operations, or to treat frequency components of the cavity as separate cavities \cite{Floquet2020}.

However, the picture may break down if additional physics beyond the basic radiation-pressure interaction are active. Notably, a Kerr-type cavity which involves thermal modulation was shown in Ref.~\cite{Kippenberg2019floquet} to result in a coupling of the processes, and to qualitatively different outcomes regarding the weights of the up- and downconverted mechanical sidebands of a thermometry tone, which compromises the interpretation of the sideband asymmetry solely as a quantum effect.

Before going to the theoretical analysis, let us discuss the sources of nonlinearities in microwave resonators at mK temperatures, effects usually associated with coupling to microscopic two-level defects (TLS) in amorphous oxides typically residing on surfaces \cite{Martinis2005TLS,Gao2008TLS,Schwab2013SiN_TLS}. The most pronounced observation is that the internal damping of the resonators can show a dependence on the applied microwave power. This power dependence is easily observable, and typically significant. Strong resonant irradiation saturates the TLS's, which leads to enhancement of the quality factor, even by several orders of magnitude, at high input microwave powers. Temperature-dependent changes due to TLS's also take place \cite{Hunklinger1976TLS}, especially of the resonant frequency, which red-shifts towards higher temperatures. 

Another mechanism that contributes to cavity nonlinearity in microwave optomechanics is a change of the BCS superconductors' surface impedance as a function of the applied power or increasing temperature, which increases the kinetic inductance of the superconducting microstrip, as a result red-shifting the cavity at high powers and elevated temperatures.
In the presented experiment, we find that this latter effect becomes relevant.



We note that in our case, the microwave resonator is a 3-dimensional cavity, which is not superconducting, and thus a question arises if there can be effects due to TLS, or breakdown of superconductivity. However, a significant amount of the electromagnetic energy resides in the aluminium thin films on the chips inside the cavity that comprise our optomechanical system.

As a summary of the above discussion, we can expect changes to both the damping rate and the resonant frequency of the microwave resonance as a function of the electromagnetic energy, i.e., basically the Kerr effect. The system is thus modeled as $H = H_0 + H_K$. The most natural choice for the origin of the Kerr effect is the energy-dependent frequency shift of the cavity, i.e.,~ $H_K = K \LL( a^\dg \RR)^2 a^2$ with the Kerr constant $K$.
%
%
%
%
%
%

For our experimental readout, we are interested in the sidebands  dropped into the output spectrum by the thermometry tone (effective coupling $G_t$), under the influence of the much stronger probe tone (effective coupling $G_p \gg G_t$). We remind the frequency spacing of the two tones; $\delta = \omega_t - \omega_p$.

We operate in the rotating frame of the probe tone, and write the cavity fields with the coherent driving fields of amplitude $A_p$ and $A_t$ for the probe and thermometry tone, respectively, and with the small fluctuation $\delta a$:
\begin{equation}
\begin{split}
     & a = A_p e^{- i \omega_p t}+A_t e^{- i (\omega_p+\delta) t} + \delta a \\
     & a^\dg = A^*_p e^{i \omega_p t}+A^*_t e^{i (\omega_p+\delta) t} + \delta a^\dg \,.
\end{split}
\end{equation}
Hereafter, we rename $\delta a \rightarrow a$.

After standard linearization of the model under the strong fields $A_p$ and $A_t$, we obtain
\begin{equation}
\label{eq:HwithKerr}
\begin{split}
     & H =  -\Delta_p a^\dg a + \omega_m b^\dg b + G_p (a^\dg + a) (b^\dg +b) + H_t +H^\m{lin}_K 
\end{split}
\end{equation}
with the thermometry part
\begin{equation}
\begin{split}
     & H_t = G_t e^{-i\delta}a^\dg (b^\dg +b) + \m{h.c.} \,.
\end{split}
\end{equation}
%

The linearized Kerr Hamiltonian $H^\m{lin}_K$ includes parametric modulation processes, and in particular a slow modulation of the cavity frequency. The latter is treated as dominant over the parametric terms, and we get
\begin{equation}
\label{eq:HKerrLin}
\begin{split}
     &  H^\m{lin}_K = 8 K A_p A_t \cos(\delta t)  a^\dg a  \,.
\end{split}
\end{equation}
We further define the effective Kerr constant $K_\m{eff}= 8 K A_p A_t$. More details are given in Appendix \ref{sec:floquet}. Based on the model in \eref{eq:HwithKerr}, we can obtain the full output spectrum, and in particular the mechanical sidebands due to the thermometry tone. In the calculation, we include the cavity internal losses and accompanying thermal cavity noise as these are significant in our experiment. In the experimentally relevant case involving also nonzero $\Delta_p$, arbitrary $\delta$, and a sideband resolution which is neither in the resolved or unresolved limit, we calculate the output spectrum numerically to compare directly to the experiment. Generally, we find that the Kerr effect introduces an artificial sideband asymmetry \cite{Kippenberg2019floquet}, which when included allows us to reach a reasonable agreement with the measurements.

\section{Experimental setup}
\subsection{Device}  
\par
\begin{figure}[htp]
    \centering
    \includegraphics[width=0.95\linewidth]{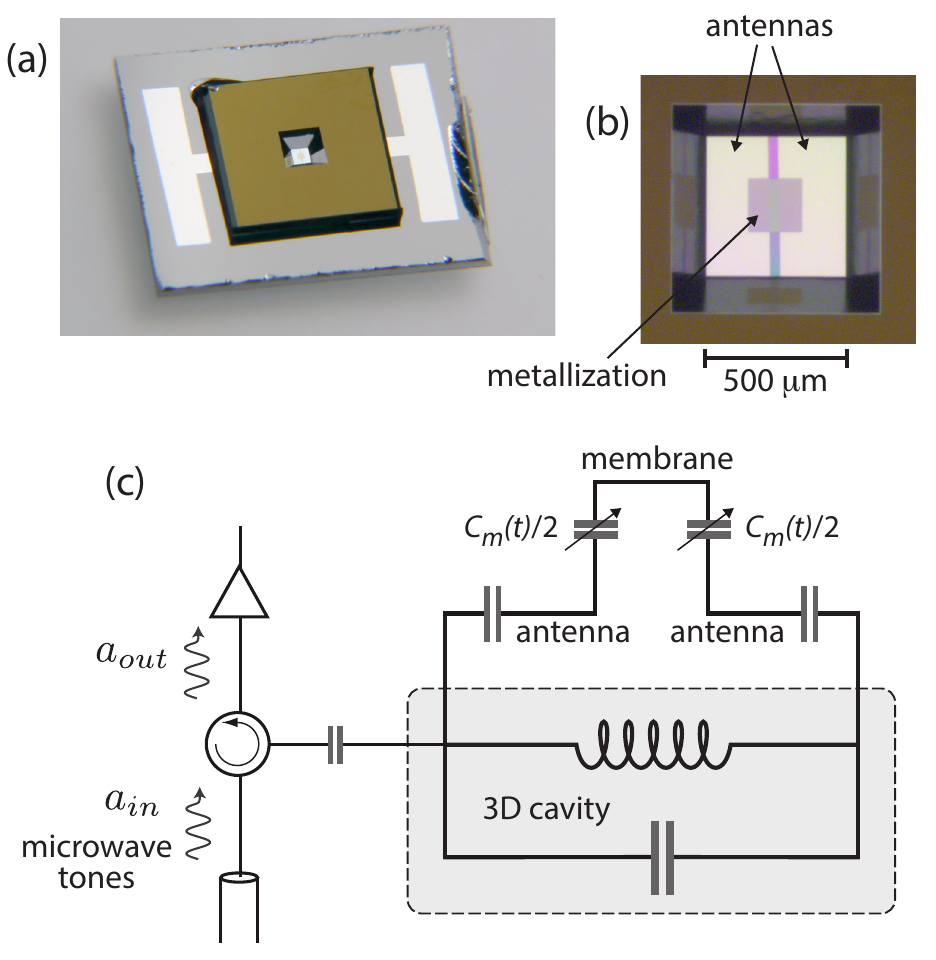}
    \caption{\emph{Device and schematics.} (a) Photograph of a representative device. The chip containing the SiN membrane appears in the center, and is flipped on top of another chip with an antenna pattern. (b) The SiN membrane is defined by back-etching silicon. The antenna pattern is visible through the membrane, with interference colors. (c) Equivalent electrical circuit of our optomechanical system. The antennas are floating with respect to the 3D cavity. Due to the split antenna design, the movable capacitance $C_m(t)$ is effectively split into two. The capacitive external coupling and a circulator are symbolically presented.}
    \label{fig:Device}
\end{figure}

 Our mechanical resonator is a 0.5 mm SiN membrane from Norcada with a thickness of 50 nm and a high tensile stress of $\sim 1$ GPa, sitting inside a $5 \times $5 mm Si frame. The membrane is metallized with 50 nm of aluminium in the form of a 200 $\mu$m square at its centre.
The metallized membrane is flipped on top of a separate patterned Si chip where a 100 nm thick aluminium antenna has been fabricated \cite{2022_SiNBAE}. Once flipped on top of the antenna, the membrane chip is glued from one of the corners using epoxy. The gap between the membrane chip and the aluminium antenna is crucial as a small gap yields a higher value of the vacuum optomechanical coupling strength $g_0$.
We estimated the gap in our device to be less than 500 nm. The chip assembly is then glued inside a 3D cavity made of annealed high-purity copper. The entire schematic is shown in \fref{fig:Device}.

The device is investigated in a dry dilution refrigerator whose base temperature is about 10 mK. Operating SiN resonators inside dry cryostats has turned out to be problematic due to large vibration noise from the pulse tube, which can strongly excite the mode to energies well above that at the thermal equilibrium at the cryostat temperature \cite{Steele3D}. In comparison, aluminum drum resonators with frequencies in the 10 MHz range and lower quality factors do not suffer significantly from the pulse tube force noise \cite{Teufel2011b,Entanglement}. The SiN resonators, however, are susceptible to pulse tube noise because the much larger quality factors enhance the sensitivity to the force noise, which already is likely larger at the lower resonant frequencies. Here we find that directly attaching the cavity box to the cryostat does not allow the mode energy to thermalize below an unacceptably high temperature $\sim 350$ mK when the pulse tube is running. In order to mitigate the problem, we have implemented an acoustic filtering described in Ref.~\cite{Gravity2021}, consisting of three 5 cm copper cubes cascaded via 3 cm diameter ring springs made of stainless steel. With the acoustic filtering, the mode energy thermalizes down to $\sim 20$ mK, and reaches the base temperature if the pulse tube is switched off.

The mechanical mode has a resonance frequency $\omega_m/ 2\pi \simeq 707.2$ kHz and has an intrinsic damping rate $\gamma/ 2\pi \simeq  9$ mHz. The resonance frequency of the cavity, including the chip inside, is $\omega_c/ 2\pi \simeq 4.554$ GHz. The internal and external damping rates of the cavity, $\kappa_i$ and $\kappa_e$, respectively, are $\kappa_i/ 2\pi \simeq 340$ kHz and $\kappa_e/ 2\pi \simeq 1.16$ MHz. The cavity is thus largely overcoupled, but not in the limit of fully dominating external coupling. This introduces some signal loss inside the cavity, which limits the feedback cooling via \eref{eq:nFB}, if $\kappa_e < \kappa$.

\subsection{Microwave scheme}   

As discussed above, the experimental setup utilizes separate tones for the purpose of probing, and for creating the feedback force, shown in \fref{fig:tones_setup}.
The probe tone is kept very close to the cavity resonance frequency, with the detuning $\Delta_p \simeq 0$ (see below), the feedback tone is above the cavity frequency at $\omega_{\m{fb}} = \omega_c + (2\pi) \cdot 32$ kHz and the thermometry tone is at the red detuning $\delta = \omega_t - \omega_p =  - (2\pi) \cdot 48$ kHz. There are several reasons behind the choice of such specific detunings. The three tones and their sideband processes are separated well enough ($\gg \gamma_\m{eff}$) in frequency that they do not overlap each other and hence can be treated independently. On the other hand, all tones are close to the cavity frequency (all detunings $\ll \kappa$) such that their dynamical backaction remains as small as possible, and they all can be simultaneously notch-filtered to remove generator phase noise. Additionally, the opposite detunings of the thermometry and feedback tones helps to cancel their dynamical backaction.

The feedback setup at room temperature follows closely that presented in Ref.~\cite{2023_feedback}. A homodyne detection is performed using the probe tone as the local oscillator.
The demodulated signal is then filtered to remove unnecessary broadband noise before being sent to the FPGA. The FPGA is a 14 bit acquisition device whose function is custom signal processing in real time.
It is programmed in such a way that it can reproduce the signal after application of band-pass filtering, time delaying, and amplification. The 
bandwidth of the digital filter, which has to be clearly larger than the maximum effective damping of the mechanics, is set at 1 kHz around the demodulated mechanical frequency.
The output of the FPGA is then sent to a phase modulator, which is driven by the feedback tone. 
The modulated feedback tone consists of two sidebands (shown in green in \fref{fig:tones_setup}). At the sample level, the sidebands mix with the tone itself to create a feedback force approximately proportional to the velocity of the mechanical oscillator. The probe, feedback and thermometry tones are combined at room temperature and notch filtered to remove generator noise before being injected into the refrigerator and sample.

The separate readout thermometry, an important part of our setup, is outside the feedback bandwidth, and is implemented out-of-loop owing to the large frequency separation between tones. Details of the thermometry calibration are given in Appendix \ref{app:analysis}.

\subsection{JTWPA characterization}   

An integral part of our experiment is a Josephson travelling wave parametric amplifier (JTWPA), described in Ref.~\cite{Perelshtein2022twpa}. This device is operated in a three-wave mixing mode, which is enabled by flux-biasing the junction chain, thus creating asymmetric nonlinearity in the Josephson potential. The device is therefore pumped at approximately twice the intended amplification frequencies. Similar to the initial designs of four-wave mixing JTWPA \cite{Siddiqi2015Amp}, the device offers several GHz of instantaneous bandwidth, with the added noise possibly close to the quantum limit of half a photon. A benefit of this design is that the relatively strong JTWPA pump tone can be easily removed by low-pass filtering with a cutoff at the upper limit $\sim 8$ GHz of the natural bandwidth of the system.

Our JTWPA is pumped at approximately 11 GHz, and its pumping power and flux bias are optimized for a large enough gain and the lowest noise. At the chosen operation point, the gain is in the range of 20 dB. This gain is sufficient such that the effective noise of the next stage amplification, typically $13$ quanta in our setup, becomes negligible.

The added noise of cryogenic microwave amplifiers is typically determined using a known tunable noise source, such as a heated matched resistor as close as possible to the input of the amplifier. We use, instead, a known noise signal arising from the microwave optomechanics itself, namely the mechanical sideband in the output spectrum, and relying on the well-established theoretical model for it. Generally speaking, we use the signal-to-noise ratio of a sideband measured at known conditions, versus the background noise floor. The details of this procedure are given in Appendix \ref{app:analysis}. We obtain the effective system added noise at the sample plane $n_{\m{add}} \simeq 2.5$ quanta. The noise of the JTWPA itself is smaller because of the inevitable, and somewhat significant losses between the sample and the amplifier. In our case, the losses are estimated to be in the range of $\sim 1$ dB, and thus the JTWPA noise is on the order 2 quanta, which is above the quantum limit because of internal losses in the device.


\begin{figure}[htp]
    \centering
    \includegraphics[width=1\linewidth]{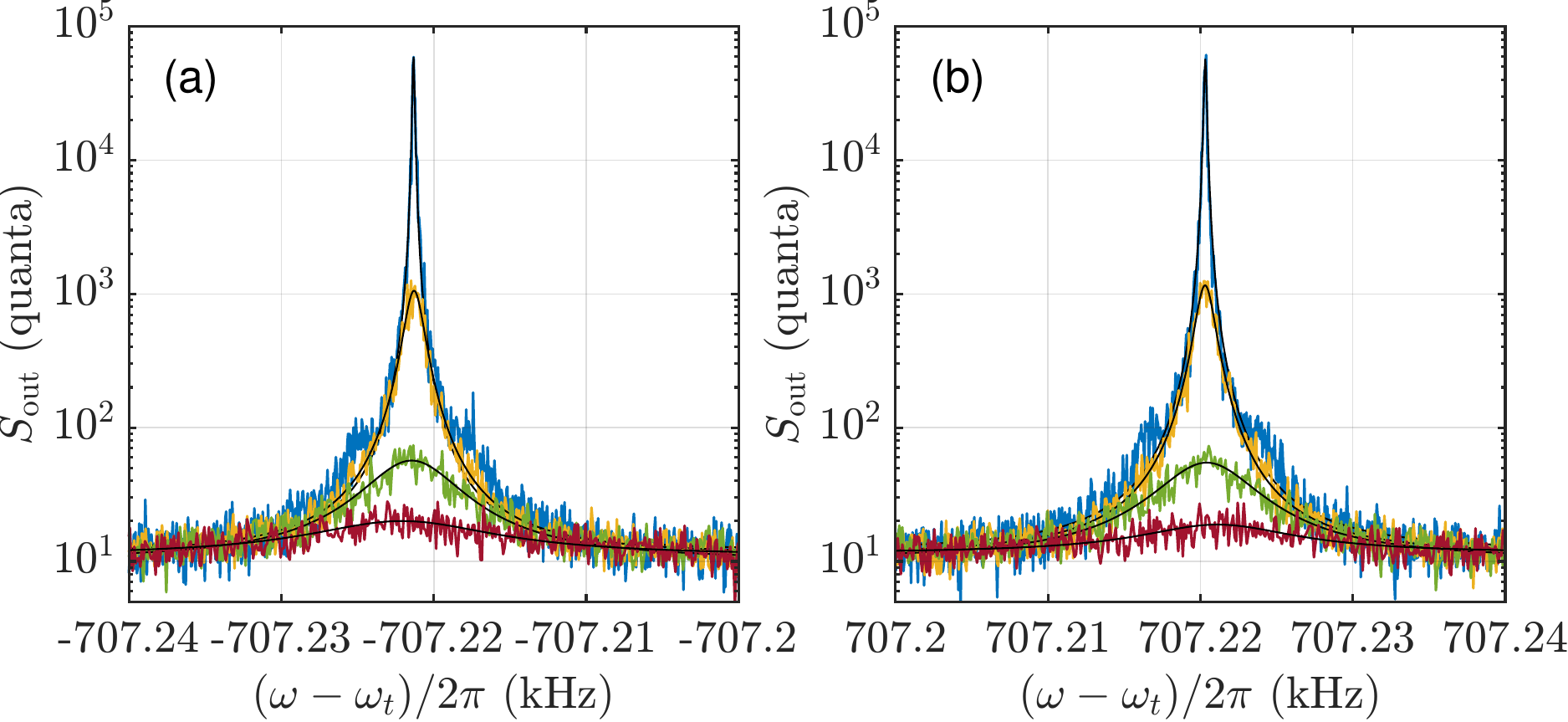}
    \caption{\emph{Thermometry signal under feedback cooling.} The thermometry sidebands showing the (a) Stokes (b) anti-Stokes peaks with relatively small feedback gain values and modest cooling. The dimensionless gain values, from top to bottom, are $[0; \,   0.23; \,  1.4; \,  3.4] $ in arbitrary units which we use consistently throughout the work. The black lines are Lorentzian fits.}
    \label{fig:spectraLowGain}
\end{figure}



\section{Results}

\subsection{Dynamical backaction}

In the ideal model case of measurement-based feedback cooling, the oscillator is damped only via the feedback. This case is satisfied with strictly zero probe detuning $\Delta_p = 0$, and vanishing powers of the thermometry and feedback tones, because all three tones will in principle contribute dynamical backaction which will either cool or heat the oscillator on top of the feedback. 

However, with the current parameters, in particular having a small intrinsic mechanical damping $\gamma$, a significant dynamical backaction will appear at nearly vanishing values of detuning of any of the tones, in particular of the strong probe tone. As an example, a mere $\sim 100$ Hz of probe detuning will contribute an optical damping equal to the intrinsic damping. Setting the detuning so accurately, in order to reach essentially $\Delta_p = 0$, is not possible based on independent calibration of the cavity frequency via linear response measurement relying on the $S_{11}$ reflection. Moreover, the cavity frequency was found to slightly drift during the experimental runs.

\begin{figure}[htp]
    \centering
    \includegraphics[width=1\linewidth]{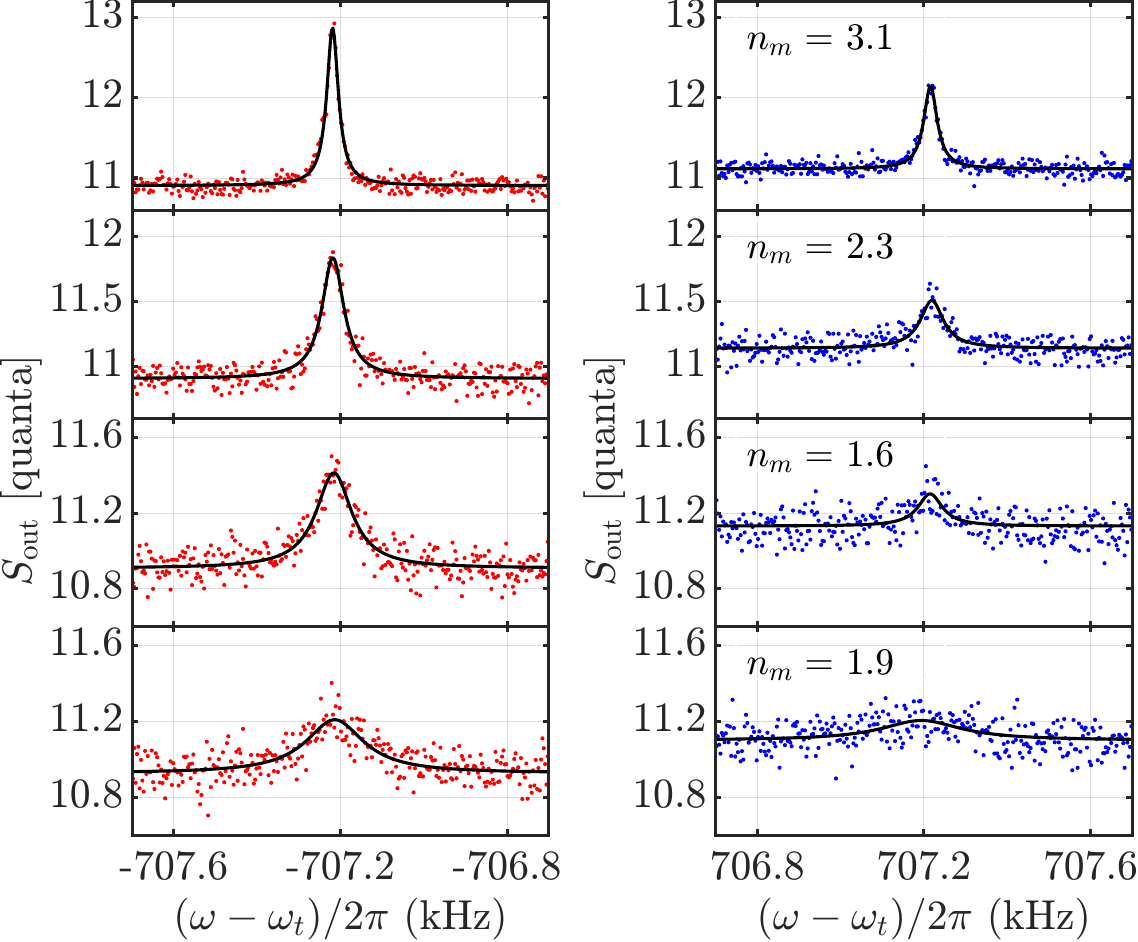}
    \caption{\emph{Thermometry signal, high feedback gains.} Left: Stokes peaks; Right: anti-Stokes peaks. The dimensionless gain values, from top to bottom, are $[8.3; \,   13.0; \, 20.4; \, 31.9] $. The obtained mechanical occupation is indicated in each panel. The solid lines are Lorentzian fits.}
    \label{fig:spectraHighGain}
\end{figure}

We aim at setting the detuning of the probe tone such that the final cooling is fully dominated by the feedback cooling. Since the thermometry and feedback tones have opposite detunings and they are not too strong, the two optical spring effects will roughly cancel. With all three tones switched on, we adjust the probe detuning such that the effective damping without feedback becomes on the order a few hundred mHz. This corresponds to $\Delta_p/2\pi \simeq -2$ kHz. While adjusting $\Delta_p$ towards a smaller absolute value results in a reduced effective damping approaching the intrinsic value, the mechanics was found to become unstable likely due to cavity drift and intermittent excursions into the blue-detuned region. We note, however, that the tones apply a significant quantum and thermal backaction heating to the oscillator, and in spite of effective damping more than an order of magnitude above the intrinsic damping, the oscillator is only cooled  by $\sim 20$ \% by the dynamical backaction associated to the three tones.

\begin{figure}[h]
    \centering
\includegraphics[width=0.85\linewidth]{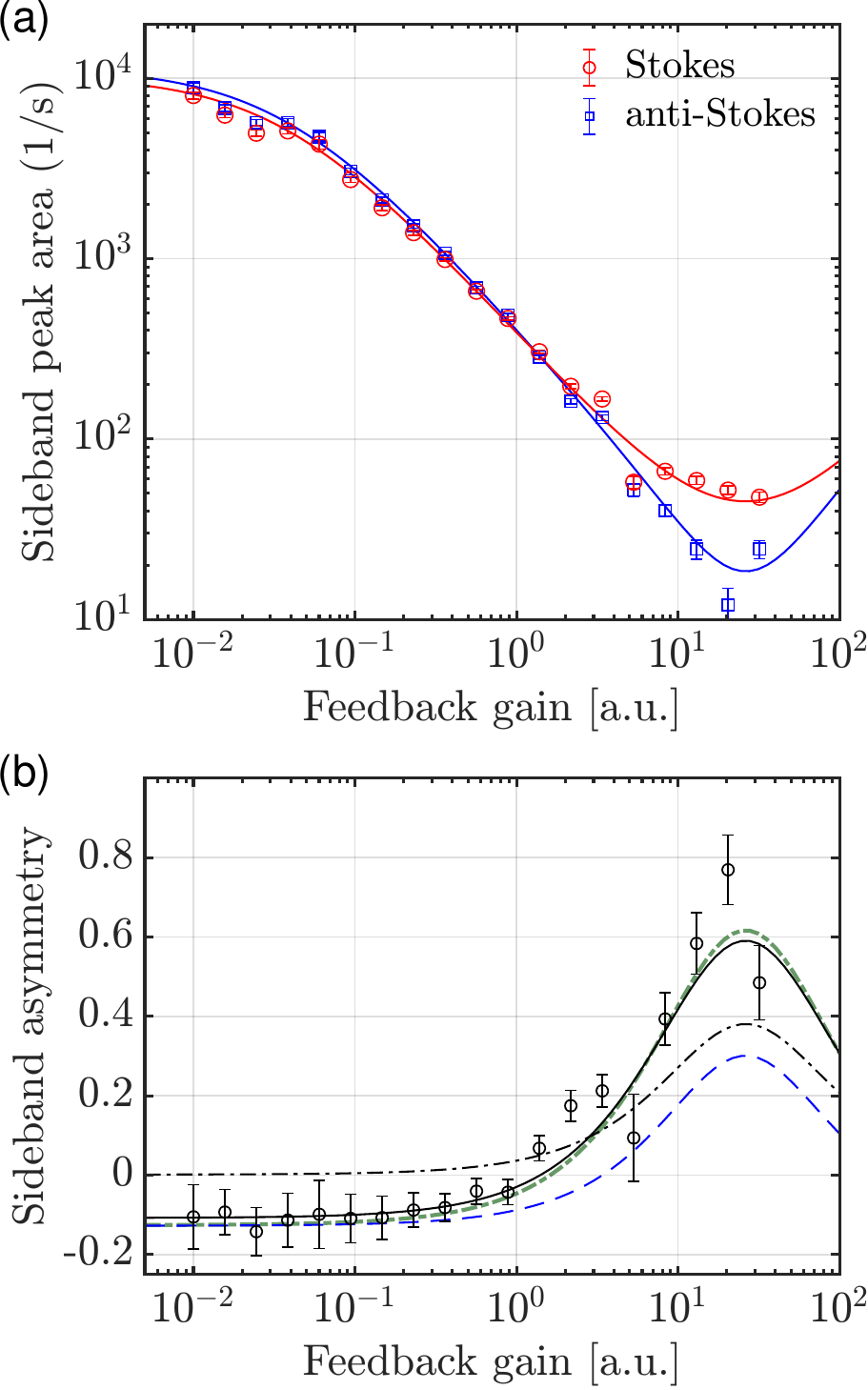}
    \caption{\emph{Sideband asymmetry in feedback cooling} (a) Sideband weights in thermometry signal, together with theoretical expectations. Effective coupling of the probe tone is $G_p/2\pi \simeq 6.32$ kHz. The solid lines colored similar to the data represent the full theoretical model including cavity Kerr effect. (b) The asymmetry, defined via the sideband weights in \eref{eq:sbasymm}. Solid line: full theoretical model including cavity Kerr effect. Short-dashed green line: theoretical model without cavity Kerr effect.  Dash-dotted line: Ideal model case; $\Delta_t = 0$, no cavity heating. Blue long-dashed line: $\Delta_t$ as in experiment, no cavity heating.}
    \label{fig:asymmetry}
\end{figure}

The noise background in the measurement is the sum of added noise and vacuum noise, i.e.,~ $n_{\m{add}} + \puoli$ quanta. Via \eref{eq:SoutCavNoise}, this information allows for determining the cavity heating taking place at strong microwave irradiation, appearing as a noise raise by a certain factor. We obtain $n_c^T \simeq 0.40 ... 0.44$ at the probe tone powers discussed in this work. This noise becomes the major limiting factor in feedback cooling.

\subsection{Feedback cooling}

To proceed with the actual feedback measurements, we calibrate the effective coupling of the probe tone via changing its detuning, and studying the effective damping with the feedback not yet applied. This gives $G_p/2\pi \simeq 6.32$ kHz for the main data which is discussed next. The phase of the feedback is optimized for minimal frequency shift and maximum damping. The two spectral peaks measured with the thermometry tone in heterodyne readout are shown in \fref{fig:spectraLowGain} at low feedback gain values. The peaks are seen to be nearly of equal height, but the right-hand-side (anti-Stokes) peak is slightly taller because of the negative detuning of the thermometry tone. At the highest feedback gains, shown in \fref{fig:spectraHighGain}, the asymmetry becomes pronounced, and also swaps sign as compared to small cooling. 

Notice that the noise floor in Figs.~\ref{fig:spectraLowGain}, \ref{fig:spectraHighGain} is approximately twice that expected from the system noise and cavity heating contribution. To reach the system noise floor in the heterodyne measurement associated to the thermometry tone, it is necessary to filter out the image frequency which introduces unnecessary noise. In the present case, the filtering was not possible due to the proximity of all the tones. Notice also that for the probe tone, the measurement is homodyne, and such a problem does not exist.

We now inspect in detail the sideband asymmetry defined in terms of the measured peak weights in the thermometry signal:
\begin{equation}
\label{eq:sbasymm}
\begin{split}
\eta = \frac{A^- - A^+}{A^-} \,.
\end{split}
\end{equation}
This is linked to the mechanical occupation via \eref{eq:nmasym}, but the important caveats discussed below \eref{eq:nmasym} have to be considered. We show in \fref{fig:asymmetry} (a) the weights of the two sidebands of the out-of-loop thermometry tone processed from the peaks shown in Figs.~\ref{fig:spectraLowGain}, \ref{fig:spectraHighGain}. At high gains above $\sim 10$, the peak areas cease decreasing because the feedback cooling looses its efficiency. The asymmetry of these peak weights is further plotted in \fref{fig:asymmetry} (b). Both panels also show the theoretical expectations based on our model, evaluated numerically. The blue long-dashed line in \fref{fig:asymmetry} (b) shows the basic model of the asymmetry, which accounts for the finite thermometry tone detuning, but no cavity heating. In this case, the asymmetry at high gain (low mechanical occupation) values is primarily due to the quantum asymmetry. Notice that the measured asymmetry is clearly higher -- due to the noise squashing by the extra cavity noise. The short-dashed green line shows the theoretical model including cavity noise. This already shows a good agreement with the measurement without any free parameters.


\begin{figure}[h]
    \centering
\includegraphics[width=0.9\linewidth]{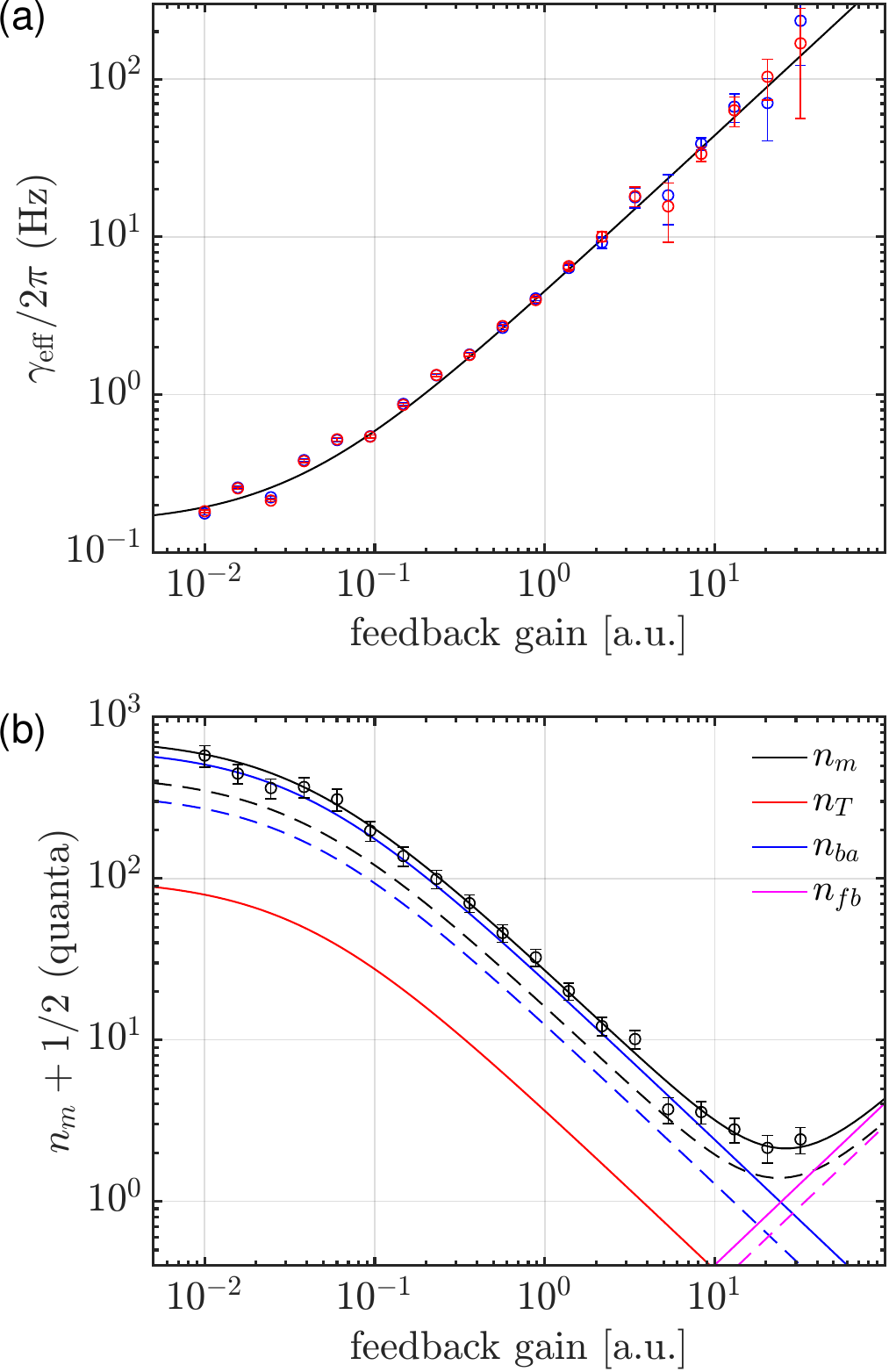}
    \caption{\emph{Feedback cooling.} (a) Effective damping of the mechanical oscillator as a function of feedback gain. The red and blue data points represent the Stokes and anti-Stokes sidebands, respectively. The solid line is a linear fit. (b) Mechanical occupation as a function of feedback gain. The theoretical curves display various contributions to the expected mechanical occupation. See Eqs.~(\ref{eq:nmFB} - \ref{eq:nFB}) for their meaning. The dashed lines colored similarly to the corresponding solid lines represent the case without cavity heating, $n_c^T = 0$.}
    \label{fig:gain_sweep}
\end{figure}

We now include the cavity Kerr effect in the theoretical model. The best fit is shown in \fref{fig:asymmetry} (b) as the solid black line. Although not very evident, this displays a better fit in the low-gain region. Without considering the cavity Kerr effect, we could not get rid of this small discrepancy even by letting the calibrated parameters float beyond their expected range. The best fit, focusing in the low-gain region where the data is less scattered, is reached at the Kerr coefficient $K_\m{eff}/2\pi \simeq 1.2$ kHz (see \eref{eq:HKerrLin}) which indicates that the cavity frequency is slowly modulated by this amount. Although the value is small as compared to e.g.~the cavity linewidth, this parameter plays a role since it is not negligible as compared to the tone detunings. The more fundamental Kerr constant $K$ is not currently accessible as we cannot  calibrate the values of the driven cavity fields $A_p$ and $A_t$ well enough.




The main result of interest, describing the ultimate feedback cooling of the oscillator, is displayed in \fref{fig:gain_sweep}. We reach the thermal occupation of the oscillator $n_m \simeq 1.6 \pm 0.4$. In the figure we also examine the limiting factors of the cooling. The dashed lines represent the expected case without cavity noise, for example, the blue dashed line is the pure quantum backaction due to the strong measurement. The thermal backaction is nearly double the quantum backaction at lower gains, and is the dominant limiting factor at the optimum gain around $\sim 20$. The cavity heating puts less severe restriction on the noise injection heating because of the suppression by the cavity susceptibility (\eref{eq:nFB}). Disregarding the cavity heating, the case which is shown with the black dashed line, ground-state cooling ($n_m < 1$) should be possible.

Based on our numerical model, we find that similar to Ref.~\cite{Kippenberg2019floquet}, the transduction of the thermometry is insensitive to the Kerr effect, which primarily distributes the same energy between the two sidebands. Based on our model, the transduction, however, is affected on the order $\sim 10$ \% by the noise squashing by cavity heating. We thus correct the thermometry signal from the measured peak areas by the expected transduction efficiency affected both by the Kerr effect and noise squashing. The corrections are also taken into account in the theoretical predictions plotted in Figs.~\ref{fig:asymmetry}, \ref{fig:gain_sweep}.

\begin{figure}[h]
    \centering
\includegraphics[width=0.8\linewidth]{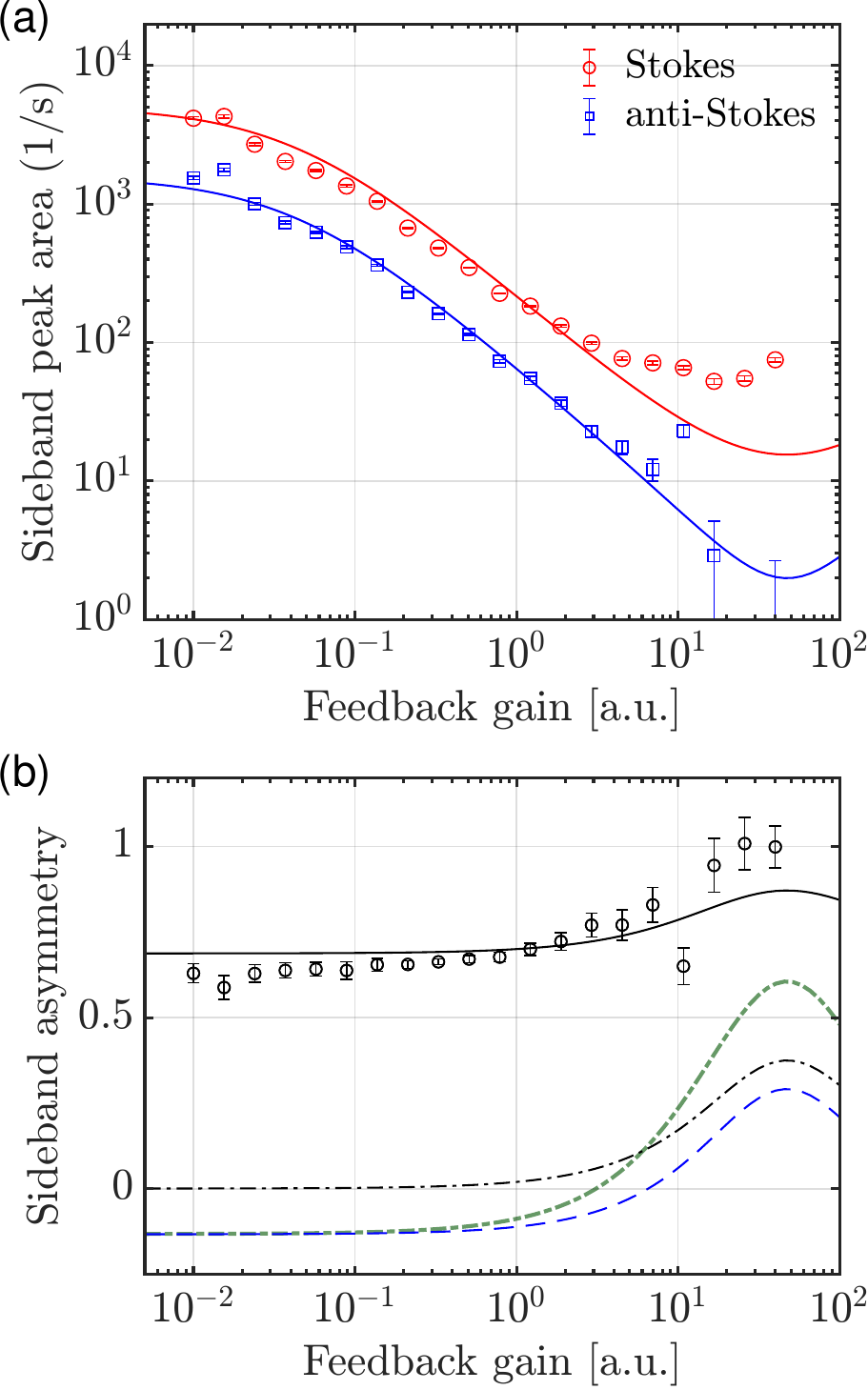}
    \caption{\emph{Artificial sideband asymmetry at a large probe power.} The graphs are similar to \fref{fig:asymmetry}, but at a large probe tone coupling $G_p/2\pi \simeq 11$ kHz.}
    \label{fig:asymmetry10}
\end{figure}

Finally, the situation with sideband asymmetry is markedly different if we increase the probe effective coupling by 5 dB, up to $G_p/2\pi \simeq 11$ kHz. This is seen in \fref{fig:asymmetry10}, which is the same as     \fref{fig:asymmetry} that was measured at lower probe power. Even in the absence of any feedback cooling the Stokes sideband is more than three times larger than the anti-Stokes sideband. This massive sideband imbalance obviously cannot be associated to the asymmetry due to the oscillator being cooled close to the ground state.

The extracted asymmetry is displayed in \fref{fig:asymmetry10} (b). It becomes reasonably well explained by the cavity Kerr effect, using the effective Kerr constant as an adjustable parameter, as shown by the solid black line in the figure. The best fit is obtained with $K_\m{eff}/2\pi \simeq 70$ kHz, a much larger value than above with the smaller probe power. In addition to the larger Kerr constant, the asymmetry in this case is also enhanced through the dynamics with the more unequal values of the couplings $G_p$ and $G_t$.

We also notice in \fref{fig:asymmetry10} (a) that the overall sideband peak area remains significantly high even at large gains $\sim 10$. The feedback cooling in this case is possible only down to $n_m \simeq 8.4$ quanta. We assume the inhibited cooling and discrepancy to the cooling model is caused by the Floquet dynamics also affecting the feedback cooling dynamics, not only the thermometry. This, however, is beyond our theoretical model.

\section{Discussion}

The largely different effective Kerr constants between the two discussed probe effective couplings, which differ by nearly two orders of magnitude, cannot be explained based on
\eref{eq:HKerrLin} at a fixed value of the original constant $K$, but the value has to be power-dependent. This is explained as a departure from BCS superconductivity via temperature-dependent kinetic inductance of the aluminum film. The kinetic inductance affects the resonance frequency of the cavity, which becomes dependent on the instantaneous temperature determined by the slow modulation of cavity energy. In our devices, cavity frequency becomes sharply temperature dependent above $\sim 250$ mK. Such very nonlinear temperature dependence is the likely reason for the strongly power-dependent $K$. Based on this, the largest probe power modulates the cavity temperature up to $350 ... 400$ mK, while the temperature modulation depth at the smaller discussed power is too small to be determined. 

\section{Conclusions}


We have shown that a feedback-controlled microwave optomechanical device involving a membrane oscillator with a high quality factor $\sim 10^8$ can be cooled close to the ground state. In order to substantially cut the noise being injected back to the device, we used a Josephson Travelling Wave Parametric Amplifier (JTWPA), which exhibits down to 2 quanta of added noise at 4.5 GHz. Technical heating of the cavity becomes a limiting factor for the feedback cooling, which enhances the measurement backaction to values clearly larger than the fundamental quantum backaction. Future improvements include reducing cavity internal losses which will likely help with cavity heating, and optimizing the JTWPA for even smaller noise. This will open the door for more sophisticated measurement-based control of mechanical motion \cite{Marquardt2008Sq,Aspelmeyer2011Telep,WoolleyBAE}, which include quantum backaction evading measurements, or mechanical teleportation.

\begin{acknowledgments} We would like to thank Laure Mercier de L\'epinay for useful discussions and Robab Najafi Jabdaraghi for parametric amplifier fabrication. We acknowledge the facilities and technical support of Otaniemi research infrastructure for Micro and Nanotechnologies (OtaNano). This work was supported by the Academy of Finland (contract 352189), and by the European Research Council (contract 101019712). The work was performed as part of the Academy of Finland Centre of Excellence program (contracts 352932, and 336810). We acknowledge funding from the European Union's Horizon 2020 research and innovation program under grant agreement 824109, the European Microkelvin Platform (EMP), and QuantERA II Programme (contract 13352189). The work at VTT was supported in part by the Research Council of Finland through Grant 321700 and through its Centres of Excellence program under Grants 352934 and 336819, in part by the EU Flagship on Quantum Technology H2020-FETFLAG-2018-03 Project under Grant 820363 OpenSuperQ, and in part by HORIZON-CL4-2022-QUANTUM-01-SGA Project under Grant 101113946 OpenSuperQPlus100.
\end{acknowledgments}



%


\vspace{1cm}

\appendix


\section{Floquet formalism for sideband mixing}
\label{sec:floquet}

The equations of motion based on the linearized Hamiltonian in \eref{eq:HwithKerr} are:
\begin{widetext}
\begin{equation}
\label{eom2tonefloq}
\begin{split}
&\dot{a} =  i \Delta a + i (\Delta_t e^{i\delta t } + \Delta_t e^{-i\delta t }) a  -  \frac{\kappa}{2} a +i G_p (b + b^\dg)  + iG_t e^{-i\delta t} (b^\dg +b) + \sqrt{\kappa_e} a_{in,e}+ \sqrt{\kappa_i} a_{in,i} \\
&\dot{a}^\dag = - i \Delta a^\dag - i (\Delta_t e^{i\delta  t} + \Delta_t e^{-i\delta t }) a^\dag  -  \frac{\kappa}{2} a^\dg - i  G_p (b + b^\dg)   - iG_t e^{i\delta t} (b^\dg +b)  + \sqrt{\kappa_e} a^\dg_{in,e}+ \sqrt{\kappa_i} a^\dg_{in,i} \\
&\dot{b} =  -i \omega_m b  -  \frac{\gamma}{2} b + i G_p( a +a^\dg) +iG_t e^{i\delta t}a + iG_t e^{-i\delta t}a^\dg + \sqrt{\gamma} b_{in} \\ 
&\dot{b}^\dg = i \omega_m b^\dg -  \frac{\gamma}{2} b^\dg - i G_p( a +a^\dg) - iG_t e^{-i\delta t}a^\dg  - iG_t e^{i\delta t}a + \sqrt{\gamma} b^\dg_{in} \,.
\end{split}
\end{equation}
\end{widetext}
Here, the input noise to the cavity arises from the external coupling ($a^\dg_{in,e}$), as well as from internal losses ($a^\dg_{in,i}$) which also describes technical heating of the cavity.

In \eref{eom2tonefloq}  we cannot ignore off-resonant processes due to a low sideband resolution $\kappa \gtrsim \omega_m$.

%
%

We start solving the system \eref{eom2tonefloq} by introducing the Floquet ansatz
\cite{Nunnenkamp2016floquet,Kippenberg2019floquet,Floquet2020}
\begin{equation}
\begin{split}
& a = \sum_n e^{i n \delta t } a_n\\  
& b = \sum_n e^{i n \delta t } b_n\\  
& a^\dg = \sum_n e^{-i n \delta t } a^\dg_n\\  
& b^\dg = \sum_n e^{-i n \delta t } b^\dg_n\,,
\end{split}
\end{equation}
where the index $n$ will be truncated over the dominant frequency components. In the end, the relevant Floquet components will be $a_0, a_{-1}, b_0$, and their Hermitian conjugates. The equations of motions for these components become
\begin{equation}
\label{eq:eom00m1}
\begin{split}
\dot{a}_0 & =  i \Delta a_0 + i e^{i \phi}  \Delta_t a_{-1}  + i  G_1  b_0 + iG_1 b^\dg_{0} \\&  - \frac{\kappa}{2}  a_0 + \sqrt{\kappa_e} a_{0in,e}+ \sqrt{\kappa_i} a_{0in,i} \\
\dot{a}^\dg_0  &=  - i \Delta a^\dg_0 - i e^{-i \phi}  \Delta_t a^\dg_{-1}  - i  G_1 b_{0} - i  G_1 b^\dg_0 \\
& - \frac{\kappa}{2} a^\dg_0 + \sqrt{\kappa_e} a^\dg_{0in,e}+ \sqrt{\kappa_i} a^\dg_{0in,i} \\
\dot{b}_0 & =  -i \omega_m  b_0  + i G_1 a_0 + iG_1 a^\dg_{0} \\
& +i G_2  a_{-1} + iG_2 a^\dg_{-1} + \sqrt{\gamma} b_{0in}  \\    
\dot{b}^\dg_0 &= i \omega_m  b^\dg_0  - i G_1 a_{0}  - i G_1 a^\dg_0 \\
& - iG_2  a^\dg_{-1}  - iG_2 a_{-1} + \sqrt{\gamma} b^\dg_{0in} \\  
-i\delta  a_{-1} + \dot{a}_{-1} & = i \Delta a_{-1} + i e^{-i \phi}  \Delta_t a_{0} +  i G_2 b_{0} + i G_2 b^\dg_{0} \\
& - \frac{\kappa}{2}  a_{-1} + \sqrt{\kappa_e} a_{(-1)in,e}+ \sqrt{\kappa_i} a_{(-1)in,i} \\
  i\delta  a^\dg_{-1} + \dot{a}^\dg_{-1}  &= -i \Delta a^\dg_{-1} - i e^{i \phi}  \Delta_t a^\dg_{0} - iG_2 b^\dg_{0} -iG_2 b_{0} \\
  & - \frac{\kappa}{2} a^\dg_{-1} + \sqrt{\kappa_e} a^\dg_{(-1)in,e}+ \sqrt{\kappa_i} a^\dg_{(-1)in,i} \\
\end{split}
\end{equation}

\begin{figure*}[ht]
    \centering
    \includegraphics[width=0.9\linewidth]{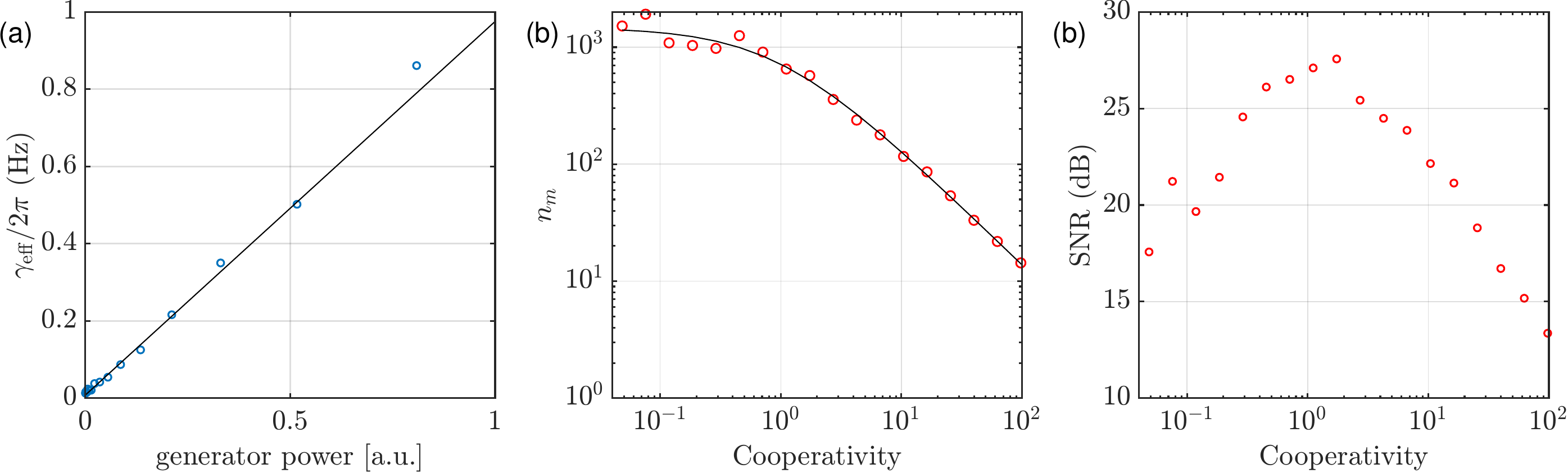}
    \caption{\emph{JTWPA noise characterization via sideband cooling.} Only the probe tone is switched on, and no feedback is applied. The power of the probe tone is varied. Refrigerator temperature was set at 50 mK, where $n_m^T \simeq 1470$ at the lowest powers studied. (a) Optical damping; (b) mechanical occupation; (c) Signal-to-Noise Ratio of the anti-Stokes peak in the output spectrum.}
    \label{fig:probeSBcoolcalib}
\end{figure*}

The quantity of interest, namely the thermometry signal occurring at $n=-1$, is written as a function of the input field as
%
\begin{equation}
\begin{split}
a_{-1}(\omega) &= M_-(\omega) a_{(-1)in,e}(\omega) +L_-(\omega) a^\dg_{(-1)in,e} (\omega) \\
&+M_{-i}(\omega) a_{(-1)in,i}(\omega)  +L_{-i}(\omega) a^\dg_{(-1)in,i} (\omega) \\
& +  Q(\omega) b_{0in}(\omega) + R(\omega) b^\dg_{0in}(\omega)   \\
a_{-1}^\dg(\omega) &= M^*_-(\omega) a^\dg_{(-1)in,e}(\omega) +L^*_-(\omega) a_{(-1)in,e} (\omega) \\
& +M^*_{-i}(\omega) a^\dg_{(-1)in,i}(\omega) +L^*_{-i}(\omega) a_{(-1)in,i} (\omega) \\
& +  Q^*(\omega) b^\dg_{0in}(\omega) + R^*(\omega) b_{0in}(\omega)  \,.
\end{split}
\end{equation}
%
The coefficients $M, L, Q, R$ pertaining to the specific input fields are solved from \eref{eq:eom00m1}, allowing for writing the output field of this component using standard procedure as
\begin{equation}
\begin{split}
& a_{(-1)out}(\omega) =  \sqrt{\kappa_e} a_{-1}(\omega) - a_{(-1)in,e}(\omega) \,
     \end{split}
\end{equation}
with the corresponding spectrum which is the final measured quantity.

\section{Data analysis}
\label{app:analysis}

\subsection{System added noise}

The noise of the Josephson Travelling Wave Parametric Amplifier (JTWPA), more precisely, the effective system noise, was characterized using the mechanical signal as a known noise source. For driving, we used the probe tone, but for this purpose, we have to switch to a heterodyne measurement. Before mixing with a separate local oscillator, the image frequency at $\omega_d - \omega_m$ was filtered away in order not to introduce doubled noise. 

We illustrate the method with the theoretical result for sideband cooling in the full resolved-sideband case. Disregarding cavity noise, the output spectrum around the anti-Stokes peak at the cavity frequency writes
\begin{equation}
\begin{split}
\label{eq:SBcoolSout}
S_{\m{out}} 
= \frac{\kappa_e}{\kappa} \frac{ \gamma_\m{opt} \gamma }{\omega^2 + \gamma^2_{\m{eff}}/4} n_m^T + n_{\m{add}} + \puoli
\,.
\end{split}
\end{equation}
The peak profile \eref{eq:SBcoolSout} is not accurate in our intermediate-sideband case, and in our actual analysis, we resort to numerical calculations. The device parameters, and the optical damping at a given generator power are easy to calibrate, and the mechanical thermal population $n_m^T$ is obtained from a standard thermal calibration. Thus, comparing the height of the Lorentzian in \eref{eq:SBcoolSout} to the noise floor, $n_{\m{add}}$ can be determined.

However, a  problem exists, which may lead to significantly under-estimating the added noise. Usually in microwave optomechanics, the mechanics suffers from “technical heating” which raises the bath temperature $n_m^T$ (and also introduces cavity noise) towards large microwave powers for reasons not fully understood. Thus, the measurement should be performed at “low” power, but it is unclear how low is needed, morever, the peak becomes possibly too weak. This issue is circumvented by utilizing the entire family of peak+background traces obtained at different powers (the usual sideband cooling experiment), and making a collective fit by using the calibrated effective coupling as a shared fit parameter. The result of this exercise is summarized in \fref{fig:probeSBcoolcalib}, where in panels (a), (b) we display the basic characterization of the sideband cooling. In panel (c), we show the signal-to-noise ratio of each peak. Although the latter does not immediately yield the added noise, the low system noise is visible as the peaks are nearly 30 dB above the noise floor. The value of added noise comes from the fit as a determined parameter.

\subsection{Thermometry}

\begin{figure}
    \centering
    \includegraphics[width=0.8\linewidth]{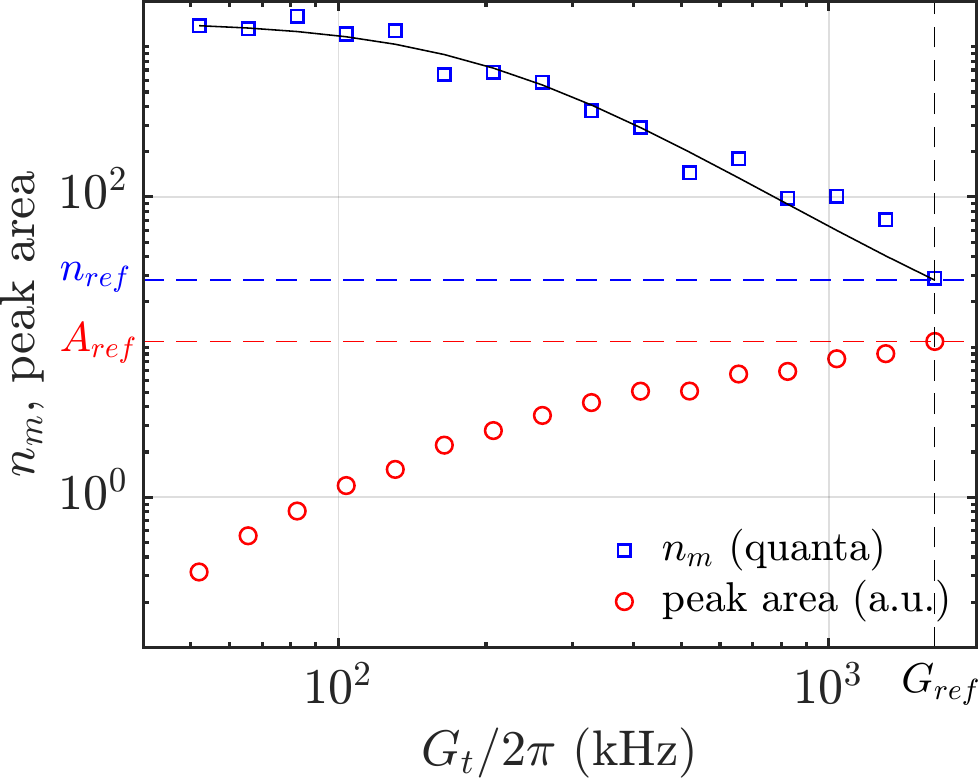}
    \caption{\emph{Thermometry calibration.} The effective coupling of the thermometry tone is varied, which causes moderate sideband cooling at the detuning $\delta/2\pi \simeq -48$ kHz.}
    \label{fig:Thermocalib}
\end{figure}

The thermometry with the separate thermometry tone allows in principle independent measurement of the mechanical energy and occupation number $n_m$. However, like we discussed above, performance of the thermometry may be compromised by the coupling of the sidebands of the pump and thermometry tones due to the cavity Kerr effect. We will take this possible complication into account in detail in the analysis. To eliminate this problem being present at the starting point, we calibrate the thermometry with only the thermometry tone present, using regular sideband cooling. 

For the thermometry calibration, we set the temperature of the refrigerator at a slightly elevated value of 50 mK to ensure the mechanical mode is reliably equilibrated at this temperature. The thermometry tone is set at the detuning $\delta/2\pi \simeq -48$ kHz, which is the same value used in the feedback measurements. Although such a small detuning is clearly sub-optimal for sideband cooling, the resulting damping, nonetheless, reaches values up to $55$ times the intrinsic damping and allows for calibrating over a large dynamic range as shown in \fref{fig:Thermocalib}. We use a rather large $G_t = G_{ref} \simeq (2\pi) \cdot 1.65$ kHz in order to recover a strong enough signal. At this value of effective coupling, we then link the phonon number $n_{ref}$ to the peak area $A_{ref}$ in the output spectrum as displayed in \fref{fig:Thermocalib}.

\end{document}